\documentclass[12pt,a4paper]{article}

\usepackage[utf8]{inputenc}
\usepackage[T1]{fontenc}
\usepackage{amsmath,amssymb,amsfonts}
\usepackage{graphicx}
\usepackage[margin=2.5cm]{geometry}
\usepackage{hyperref}
\usepackage{booktabs}
\usepackage{caption}
\usepackage{subcaption}
\usepackage{float}
\usepackage{xcolor}
\usepackage{authblk}
\usepackage{bm}
\graphicspath{{figures/}}
\usepackage[backend=bibtex, style=numeric-comp, sorting=none]{biblatex}
\newcommand{\bru}{\textsc{bru}}
\newcommand{\vect}[1]{\bm{#1}}
\newcommand{\mat}[1]{\mathbf{#1}}

\title{Blobel's Regularized Unfolding: \\
  Eigenmode Decomposition and Automatic Smoothing \\
  for Inverse Problems in Particle Physics}

\author{Vincent Alexander Croft}
\affil[1]{
    Nikhef -  Dutch National Institute for Subatomic Physics
}

\date{\today}

\begin{document}
\maketitle

\begin{abstract}
This document presents a self-contained treatment of regularized unfolding based on  cubic B-spline representations and eigenmode filtering, following the original formulation by Blobel and direct translation of the original implementation in Fortran into a modern format.  The method, which has been called by several names under its various historical representations, is named here as Blobel's 
Regularized Unfolding (\bru).
This method differs from conventional histogram-based unfolding approaches in that the true distribution is represented as a smooth function parametrised by spline coefficients, and the regularization  operates through an eigenmode decomposition of the curvature penalty relative to the statistical precision. 
This document describes the mathematical structure of the method, the mechanism by which the regularization strength is determined automatically from the data, and provides a detailed comparison with standard methods including Tikhonov regularization based methods, Richardson-Lucy iteration, and naive matrix inversion.
\end{abstract}

% =====================================================================
\section{Introduction}
\label{sec:intro}
% =====================================================================

The extraction of physical distributions from measured data in particle physics experiments requires the solution of an inverse problem. 
Detector effects such as finite resolution, acceptance losses, and bin migration distort the true underlying spectrum, and recovering that spectrum from the observed data constitutes an ill-posed problem in that small perturbations in the measured data can produce arbitrarily large fluctuations in the inferred solution.

The standard approaches to controlling this instability fall into two broad categories.
Iterative methods, exemplified by the  Richardson-Lucy algorithm~\cite{richardson1972,lucy1974} and later promoted in particle physics by D'Agostini~\cite{dagostini1995}, generate a sequence of approximations that converge toward the unregularized inverse but are truncated before the noise-amplifying modes dominate.
Matrix regularization methods, typified by Tikhonov regularization~\cite{tikhonov1963} and its implementation in TUnfold~\cite{schmitt2012}, add a penalty term to the least-squares objective that suppresses oscillatory solutions.
Both approaches introduce a tuning parameter (the number of iterations or the regularization strength) whose selection can significantly affect the result.

Blobel proposed an alternative framework~\cite{blobel1985,blobel2002} that departs from the histogram-based paradigm in a fundamental way.  
Rather than unfolding bin-by-bin, the true distribution is represented as a smooth function expressed in a B-spline basis, and the response of the detector is encoded as a matrix mapping spline coefficients to 
predicted bin counts.
The key innovation lies in the treatment of regularization: through a simultaneous diagonalization of the curvature matrix and the Fisher information matrix, the problem decomposes into independent eigenmodes whose signal-to-noise ratios can be assessed individually.
The regularization strength is then chosen by requiring that the suppressed modes contribute a chi-squared increment consistent with their statistical expectation, eliminating the need for external tuning.

This paper provides a modern, self-contained exposition of Blobel's method, which we will refer to as \textbf{Blobel's Regularized Unfolding} (\bru{}).
Section~\ref{sec:inverse} formulates the inverse problem and introduces the B-spline representation.
Section~\ref{sec:eigenmode} develops the eigenmode decomposition that underlies the regularization.
Section~\ref{sec:auto_tau} describes the automatic selection of the regularization parameter.  Section~\ref{sec:numerical} presents numerical studies comparing \bru{} with Tikhonov regularization, Richardson-Lucy iteration, and naive inversion on two benchmark distributions: a double-peaked spectrum and a steeply falling power-law spectrum.
Section~\ref{sec:discussion} discusses the results and outlines directions for further development.  
Section~\ref{sec:interpretation} addresses the interpretation of unfolded results in downstream inference, arguing that the eigenmode decomposition of \bru{} provides an unusually transparent basis for this purpose with a conclusion presented in section~\ref{sec:conclusions}.

% =====================================================================
\section{The inverse problem and spline representation}
\label{sec:inverse}
% =====================================================================

We consider a measurement in which an unknown true distribution $f(x)$ of a variable $x$ is observed through a detector whose response is characterized by a conditional probability density $R(y|x)$, so that the expected number of events in measured bin $i$ is
\begin{equation}
  \mu_i = \int R(y_i | x)\, f(x)\, \mathrm{d}x \,,
  \label{eq:forward}
\end{equation}
where the integral of $R(y_i|x)$ over the $i$-th measured bin is implied.
The observed data $\{n_i\}$ are Poisson-distributed with expectations $\{\mu_i\}$.
The goal is to recover $f(x)$ from the observed counts.

In conventional approaches the true distribution is discretized into histogram bins, producing a finite-dimensional linear system $\vect{\mu} = \mat{A}\,\vect{f}$, where $\mat{A}$ is the response (or migration) matrix. 
The difficulty is that $\mat{A}$ is typically rank-deficient or ill-conditioned, so that the naive least-squares solution is dominated by statistical noise amplified through the small singular values of $\mat{A}$.

\bru{} replaces the histogram discretization with a smooth representation in terms of cubic B-splines\cite{deboor1978}.
The true distribution is written as
\begin{equation}
  f(x) = \sum_{j=1}^{p} c_j\, B_j(x) \,,
  \label{eq:spline}
\end{equation}
where $B_j(x)$ are cubic B-spline basis functions defined on a uniform knot sequence with $p$ knots spanning the range $[x_{\mathrm{low}}, x_{\mathrm{high}}]$, and the coefficients $\{c_j\}$ are the parameters to be determined. 
The number of spline knots $p$ is typically chosen to be somewhat larger than the number of resolvable features in the distribution, since the regularization will effectively suppress the coefficients associated with high-frequency oscillations.

Substituting Eq.~(\ref{eq:spline}) into Eq.~(\ref{eq:forward}) yields a linear relationship between the spline coefficients and the expected counts:
\begin{equation}
  \mu_i = \sum_{j=1}^{p} R_{ij}\, c_j \,,
  \label{eq:linear}
\end{equation}
where $R_{ij} = \int R(y_i | x)\, B_j(x)\, \mathrm{d}x$ is the response matrix in the spline basis.  In practice, $R_{ij}$ is constructed from a Monte Carlo simulation by processing simulated events through the detector response and accumulating the resulting weights in the spline basis representation.

\begin{figure}[tb]
  \centering
  \includegraphics[width=\textwidth]{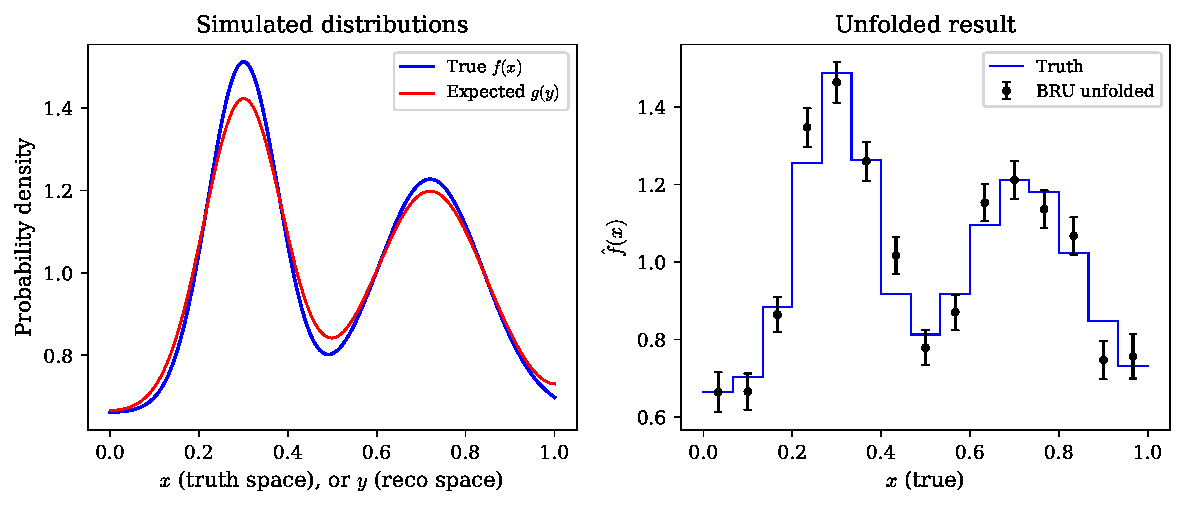}
  \caption{Illustration of the inverse problem.  \textbf{Left:} The 
    true distribution $f(x)$, a double-peaked spectrum and the measured distribution $g(y)$ demonstrating the effect of the detector 
    smearing.  \textbf{Right:} The unfolded result from \bru{} 
    (black points with error bars) compared with the truth (blue 
    histogram).}
  \label{fig:inverse_problem}
\end{figure}

Figure~\ref{fig:inverse_problem} illustrates the various stages of the problem: the true distribution $f(x)$ is smeared by the detector response to produce the measured spectrum $g(y)$, and the goal of unfolding is to recover an estimate $\hat{f}(x)$ of the original distribution from the measured data.
This construction therefore involves 4 distributions.
The true distribution and measured spectrum are typically modelled in advance by using labelled data for example from Monte Carlo simulation. 
The actual unfolding will be performed on the data measured by your detector, to produce an unfolded distribution realised in the truth dimension. 

It should be noted here that the truth and reconstructed dimensions need not be the same.
For example you might measure the energy of an object hitting a target and wish to infer its mass.
Similarly, the unfolded spectrum of a measured distribution represented as a histogram does not need to have the same binning as the measured distribution itself. 
Blobel refers to this minimally constrained system as the "inverse crime"\cite{blobel2010lectures}\footnote{This originates in other documents \cite{hansen2010,kaipio2007} describing the use of the same computational grid to simulate the data and to solve the inverse problem. This will typically lead to the discretisation errors in the forward model and the reconstruction cancelling exactly, giving unrealistically good results. For example, constructing an unfolding problem with identical bins in truth and measured dimensions gives a square response matrix, making the inversion artificially well-conditioned.}, and instead advocated for twice the resolution of measured spectra to that inferred from it, which is the standard used in this work. 

% =====================================================================
\section{Eigenmode decomposition and regularization}
\label{sec:eigenmode}
% =====================================================================

The unregularized fit of the spline coefficients to the data minimizes 
the weighted least-squares objective
\begin{equation}
  S(\vect{c}) = \sum_{i=1}^{m} \frac{(n_i - \mu_i(\vect{c}))^2}{n_i}
  = (\vect{n} - \mat{R}\,\vect{c})^T \mat{W}\, 
    (\vect{n} - \mat{R}\,\vect{c}) \,,
  \label{eq:chi2}
\end{equation}
where $\mat{W} = \mathrm{diag}(1/n_i)$ is the weight matrix.  The solution of this problem can be given in closed form as:
\begin{equation}
\hat{\vect{c}} = (\mat{R}^T \mat{W} 
\mat{R})^{-1} \mat{R}^T \mat{W}\, \vect{n}    
\end{equation} 
To understand why this solution is unstable it can be informative to introduce the Fisher information matrix as $\mat{F} = \mat{R}^T \mat{W} \mat{R}$. This matrix governs the information content that can be used to constrain the free parameters of the system. In the unfolding case, $\mat{F}$ can have eigenvalues spanning many orders of magnitude.

Regularization introduces a penalty on the roughness of the solution.  
A natural measure of roughness for a spline function is the integrated 
squared second derivative:
\begin{equation}
  P(\vect{c}) = \int_{x_{\mathrm{low}}}^{x_{\mathrm{high}}} 
  \left[ f''(x) \right]^2 \mathrm{d}x 
  = \vect{c}^T \mat{C}\, \vect{c} \,,
  \label{eq:penalty}
\end{equation}
where $\mat{C}$ is the curvature matrix whose elements are $C_{jk} = \int B_j''(x)\, B_k''(x)\, \mathrm{d}x$.  Because the B-spline basis functions have local support, $\mat{C}$ is a banded matrix that can be computed analytically.

The regularized objective is then
\begin{equation}
  Q(\vect{c}; \tau) = S(\vect{c}) + \tau\, P(\vect{c}) 
  = (\vect{n} - \mat{R}\vect{c})^T \mat{W}\, 
    (\vect{n} - \mat{R}\vect{c}) + \tau\, \vect{c}^T \mat{C}\, \vect{c} \,,
  \label{eq:regularized}
\end{equation}
where $\tau \geq 0$ is the regularization parameter controlling the trade-off between fidelity to the data and smoothness of the solution.
The regularized solution is
\begin{equation}
  \hat{\vect{c}}_\tau = (\mat{F} + \tau\,\mat{C})^{-1} 
  \mat{R}^T \mat{W}\, \vect{n} \,.
  \label{eq:reg_solution}
\end{equation}

The central insight of Blobel's approach is that the interplay between the Fisher information $\mat{F}$ and the curvature penalty $\mat{C}$ is best understood through a simultaneous diagonalization.
Because $\mat{F}$ is positive semi-definite and $\mat{C}$ is positive semi-definite, there exists a transformation $\mat{U}$ such that $\mat{U}^T \mat{F}\, \mat{U} = \mat{I}$ (the identity) and $\mat{U}^T \mat{C}\, \mat{U} = \mat{D} = \mathrm{diag}(d_1, d_2, 
\ldots, d_p)$, where the generalized eigenvalues $d_k$ are ordered as $d_1 \leq d_2 \leq \cdots \leq d_p$.

In the transformed basis $\vect{a} = \mat{U}^{-1} \vect{c}$, the 
regularized objective decomposes into a sum of independent 
one-dimensional problems:
\begin{equation}
  Q(\vect{a}; \tau) = \sum_{k=1}^{p} \left[ 
    (a_k - \hat{a}_k)^2 + \tau\, d_k\, a_k^2 \right] + \mathrm{const.}
  \label{eq:decoupled}
\end{equation}
where $\hat{a}_k$ is the unregularized amplitude in mode $k$.  The 
regularized amplitude is
\begin{equation}
  \hat{a}_k^{(\tau)} = \frac{\hat{a}_k}{1 + \tau\, d_k} \,,
  \label{eq:filtered}
\end{equation}
and the factor $h_k = 1/(1 + \tau\, d_k)$ is called the \emph{filter 
factor} for mode $k$.  Modes with $\tau\, d_k \ll 1$ are essentially 
unmodified by the regularization, while modes with $\tau\, d_k \gg 1$ 
are strongly suppressed.

The eigenvalue $d_k$ measures the curvature cost of exciting mode $k$: 
low-$k$ modes describe smooth, large-scale features of the distribution 
and have small eigenvalues, while high-$k$ modes correspond to rapidly 
oscillating patterns with large curvature and therefore large $d_k$.  
The regularization thus acts as a low-pass filter in the eigenmode 
space, preserving the well-determined smooth components while 
suppressing the noise-dominated oscillatory components.

\begin{figure}[tb]
  \centering
  \includegraphics[width=\textwidth]{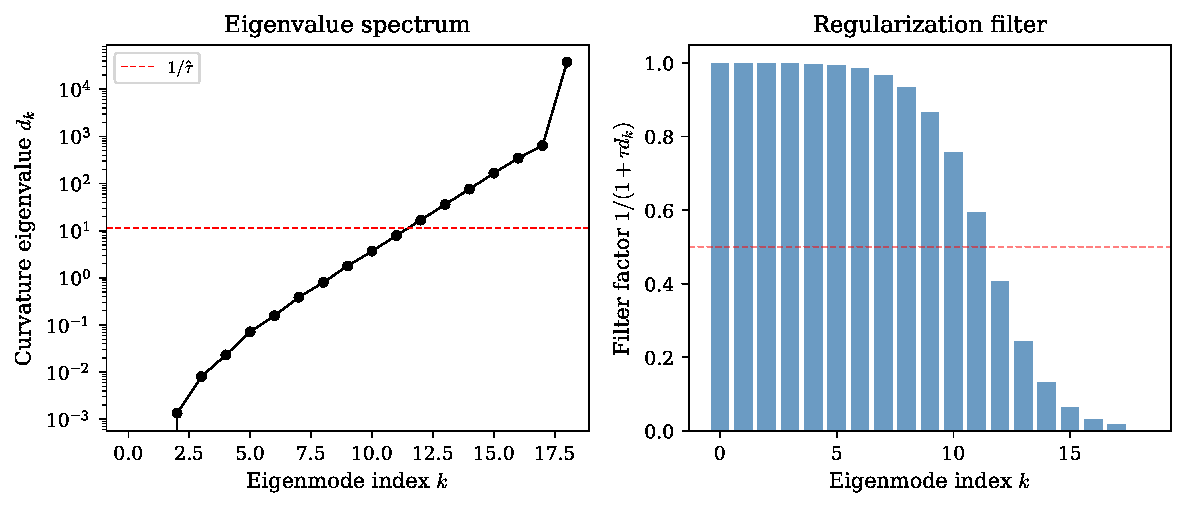}
\caption{Left: Eigenvalue spectrum of the curvature matrix in the
Fisher-normalised basis.  The dashed line indicates $1/\hat\tau$,
where $\hat\tau$ is the regularization parameter selected
automatically by the criterion of Eq.~(12); for this example
$1/\hat\tau = 12$.  Right: Filter factors $h_k = 1/(1+\hat\tau\, d_k)$.
Modes with $h_k \approx 1$ are retained; modes with $h_k \approx 0$
are suppressed.}
  \label{fig:eigenvalues}
\end{figure}

Figure~\ref{fig:eigenvalues} shows the eigenvalue spectrum and the 
corresponding filter factors for a representative unfolding of the 
double-peaked test distribution.  The eigenvalues span several orders 
of magnitude, reflecting the wide range of length scales present in 
the problem.  The filter factors transition smoothly from unity (no 
suppression) for the low-index modes to near zero for the highest-index 
modes, with the transition region determined by the regularization 
parameter $\tau$.

\begin{figure}[htb]
  \centering
  \includegraphics[width=0.65\textwidth]{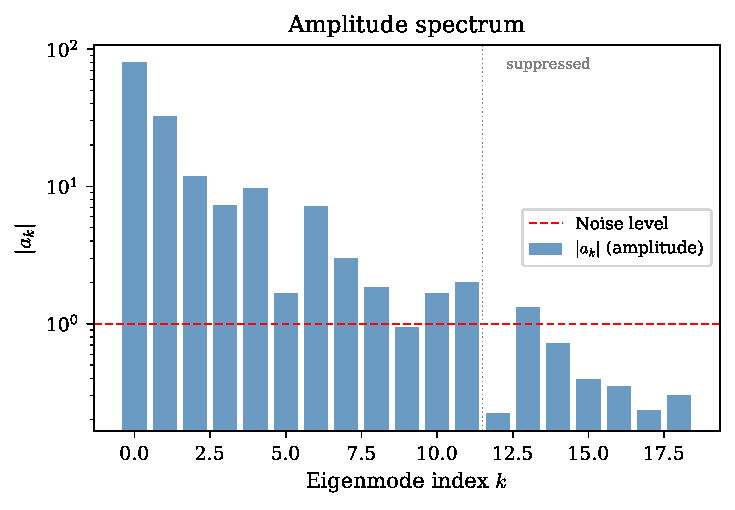}
  \caption{Amplitude spectrum of the unregularized solution.  
    The horizontal dashed line marks the noise level.  Modes to 
    the right of the vertical dotted line are suppressed by the 
    regularization.}
  \label{fig:amplitudes}
\end{figure}

Figure~\ref{fig:amplitudes} displays the amplitude spectrum 
$|\hat{a}_k|$ for the same example.  The low-index amplitudes carry 
the physical signal and are well above the noise level (indicated by 
unity on the normalized scale), while the high-index amplitudes have 
magnitudes consistent with statistical noise.  The regularization 
correctly identifies and suppresses precisely those modes where the 
signal has been lost in the noise.

\begin{figure}[htb]
  \centering
  \includegraphics[width=\textwidth]{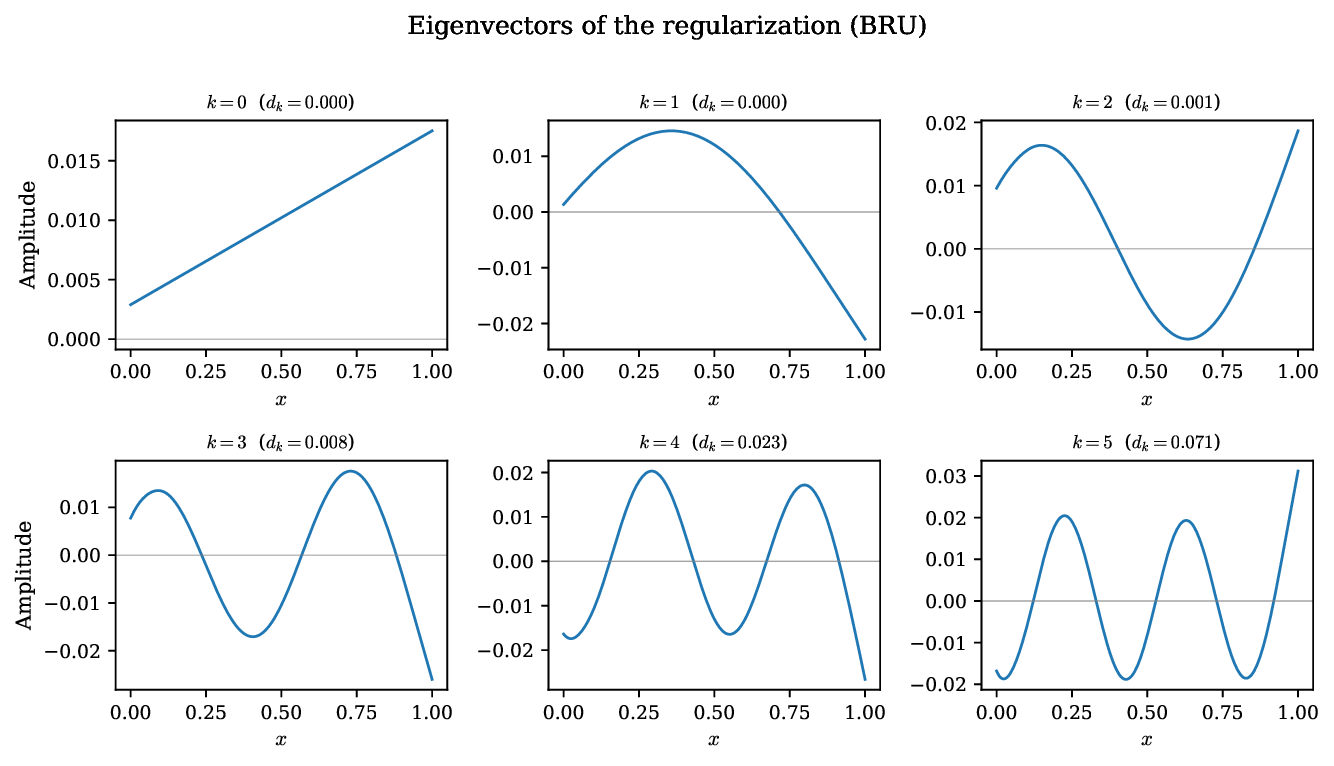}
  \caption{The first six eigenvectors of the regularization, evaluated 
    in the B-spline basis.  Low-index modes describe smooth variations; 
    high-index modes capture progressively finer structure.  The 
    eigenvalue $d_k$ is indicated in each panel title.}
  \label{fig:eigenvectors}
\end{figure}

The eigenvectors themselves provide physical insight into what each 
mode represents.  Figure~\ref{fig:eigenvectors} shows the first six 
eigenvectors evaluated in the B-spline basis.  The lowest modes 
represent overall normalization and broad tilts, while successive 
modes capture features at progressively finer scales.  The 
regularization preserves those modes that are well-constrained by 
the data while filtering out the high-frequency oscillations that 
the measurement cannot resolve.

% =====================================================================
\section{Automatic selection of the regularization parameter}
\label{sec:auto_tau}
% =====================================================================

The regularization parameter $\tau$ controls the boundary between the modes that are retained and those that are suppressed.  Choosing $\tau$ too small leaves the solution vulnerable to noise amplification; choosing $\tau$ too large over-smooths the result and introduces bias by removing genuine features of the distribution.

The consequences of this choice extend beyond the shape of the point estimate.  As demonstrated in the comparative study of Ref.~\cite{brenner2020}, optimizing $\tau$ solely on mean-squared error produces substantial undercoverage for every regularised method examined, because the MSE-optimal solution tolerates a level of bias that the reported uncertainties do not reflect.
The appropriate figure of merit is therefore not the MSE alone but the frequentist coverage of the reported confidence intervals.  Following Kuusela~\cite{kuusela2016} and the procedure adopted in Ref.~\cite{brenner2020}, the appropriate metric for the ideal choice of regularisation is the MSE performance combined with the analytical per-bin coverage for a Gaussian interval centred at $\hat\mu_i$ with reported standard deviation $\sigma_{\hat\mu_i}$ is
\begin{equation}
  C(\gamma_i)
    = \Phi(\gamma_i + z) - \Phi(\gamma_i - z) \,,
  \label{eq:coverage}
\end{equation}
where $\gamma_i = b_i / \sigma_{\hat\mu_i}$ is the bias-to-standard-error ratio in bin $i$. $b_i$ is the bias estimated as the mean pull across pseudo-experiments and $z = \Phi^{-1}(0.8415)$ is the quantile corresponding to 68.3\% nominal coverage.
At $\gamma_i = 0$ this returns exactly $68.3\%$; it decreases monotonically as $|\gamma_i|$ grows.
A method whose reported uncertainties fully account for the regularization bias will achieve coverage at or near the nominal level; one that underestimates the bias will undercover.
This diagnostic is used throughout Section~5 to compare the methods studied here.

Several approaches to selecting $\tau$ have been proposed in the literature.
Notable examples include the L-curve method~\cite{hansenLcurve} and the global correlation coefficient (GCC)~\cite{schmitt2012}, the minimisation of which selects the $\tau$ at which the covariance structure is most nearly diagonal.
Both approaches, however, require an external scan over candidate values of $\tau$ and neither directly targets the coverage of the resulting confidence intervals.

\bru{} determines $\tau$ from an internal consistency condition.
The key observation is that the chi-squared contribution of the suppressed modes has a well-defined statistical expectation.
In the absence of regularization, each mode contributes on average one unit of chi-squared.
When regularization is applied, the suppressed modes contribute
\begin{equation}
  \Delta \chi^2_{\mathrm{supp}} = \sum_{k:\, \tau d_k > 1} 
  \left( \hat{a}_k^{(\tau)} - \hat{a}_k \right)^2 
  = \sum_{k:\, \tau d_k > 1} \hat{a}_k^2 
  \left( \frac{\tau d_k}{1 + \tau d_k} \right)^2 \,.
  \label{eq:deltachi2}
\end{equation}

If the regularization is correctly calibrated, the suppressed modes should contain only noise, and their chi-squared contribution should be consistent with the number of suppressed degrees of freedom.
Blobel's criterion requires that
\begin{equation}
  \Delta \chi^2_{\mathrm{supp}} \approx n_{\mathrm{supp}} \,,
  \label{eq:criterion}
\end{equation}
where $n_{\mathrm{supp}}$ is the effective number of suppressed modes.  
In practice, $\tau$ is adjusted iteratively until this condition is satisfied to a prescribed tolerance.

The effective number of degrees of freedom in the regularized fit is $\nu_{\mathrm{eff}} = \sum_k h_k$, which counts the modes weighted by their filter factors.
An alternative formulation of the automatic selection criterion is that the total chi-squared per degree of freedom should be close to unity.
The two formulations are equivalent in the limit where the filter factors are close to zero or one, which is the regime where the regularization has a clean interpretation as mode selection.

The strength of this approach is that no external tuning is required: the data themselves determine how much smoothing is appropriate.
This is particularly valuable in the context of particle physics measurements, where the analyst should not need to impose subjective choices on the regularization strength.

% =====================================================================
\section{Numerical studies}
\label{sec:numerical}
% =====================================================================

The performance of \bru{} is evaluated using two benchmark distributions that represent qualitatively different challenges for unfolding algorithms.
The first is a double-peaked distribution with a flat background, which tests the ability to resolve localized features.  
The second is a steeply falling power-law spectrum with a superimposed bump, which tests performance when the dynamic range of the distribution spans an order of magnitude.
In both cases, the detector response is modelled as Gaussian smearing with resolution $\sigma = 0.04$ over the unit interval $[0, 1]$.
The measured distribution is binned into 30 bins, the evaluation grid uses 15 bins, and each pseudo-experiment contains 8000 events.

Four unfolding methods are compared. 
\bru{} uses 20 cubic B-spline knots with automatic regularization as described in the preceding sections.
Tikhonov regularization employs a second-difference curvature penalty with the regularization strength selected by minimizing the mean global correlation coefficient over a logarithmic scan of $\tau \in [10^{-10}, 10^{-2}]$, matching the recommended behaviour of TUnfold~\cite{schmitt2012}.
The Richardson-Lucy algorithm (also known as either Bayes or iterative Bayes in RooUnfold\cite{adye2011} or as D'Agostini's algorithm) is run for 4 iterations. 
This number of iterations is a typical choice in LHC analyses and is used here in order to demonstrate the severe undercoverage this default number provides.
In practice the author would like to remind all unfolding practitioners that the default parameter in a piece of software such as RooUnfold may not be suitable for your analysis. 
Naive matrix inversion uses the Moore-Penrose pseudoinverse of the response matrix without any regularization.

\subsection{Double-peaked distribution}
\label{sec:double_peaked}

The true distribution for this test problem is
\begin{equation}
  f(x) \propto 0.7 + 0.9\, e^{-\frac{1}{2}\left(\frac{x - 0.3}{0.08}\right)^2} 
  + 0.6\, e^{-\frac{1}{2}\left(\frac{x - 0.72}{0.12}\right)^2} \,,
  \label{eq:double_peaked}
\end{equation}
normalized to unit integral.
The two Gaussian peaks sit atop a flat continuum and have widths comparable to the detector resolution, making the narrower peak at $x = 0.3$ particularly challenging to resolve.

\begin{figure}[htb]
  \centering
  \includegraphics[width=0.75\textwidth]{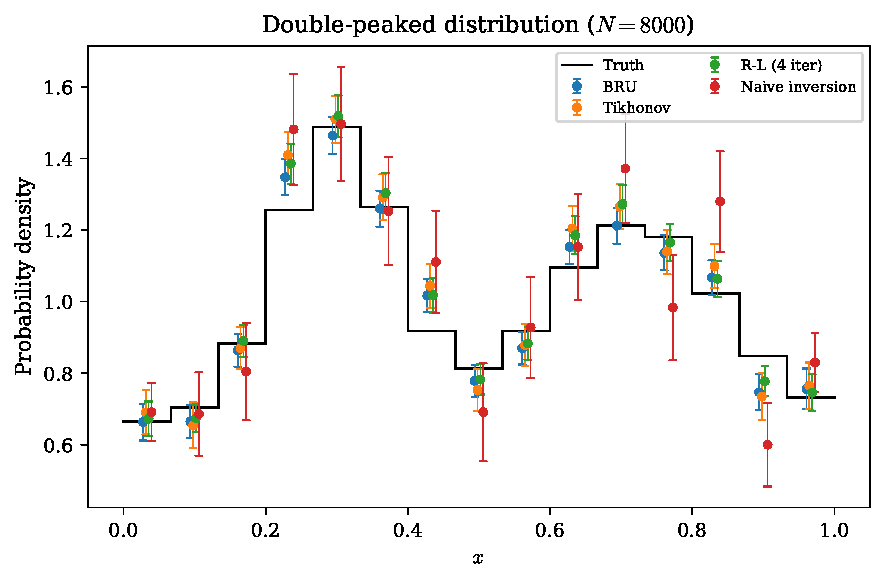}
  \caption{Single pseudo-experiment for the double-peaked distribution.  
    The solid histogram shows the truth; points with error bars show 
    the unfolded results from each method.}
  \label{fig:singletoy_dp}
\end{figure}

Figure~\ref{fig:singletoy_dp} shows the result of a single pseudo-experiment.
\bru{} tracks the truth closely across both peaks and in the inter-peak valley, with error bars that reflect the local statistical precision.
Tikhonov regularization shows a slight positive bias in the peak regions due to over-smoothing, while Richardson-Lucy iteration produces noticeable oscillations around the narrower peak.
Naive inversion exhibits the expected large fluctuations associated with noise amplification.

To assess the statistical properties of each method, 1,000 pseudo-experiments 
are generated with independent Poisson fluctuations.  For each toy and 
each evaluation bin, the pull is defined as 
$p_i = (\hat{f}_i - f_i^{\mathrm{true}}) / \sigma_i$, where $\sigma_i$ 
is the reported uncertainty.  A well-calibrated method should produce 
pulls that follow a standard normal distribution.

\begin{figure}[htb]
  \centering
  \includegraphics[width=\textwidth]{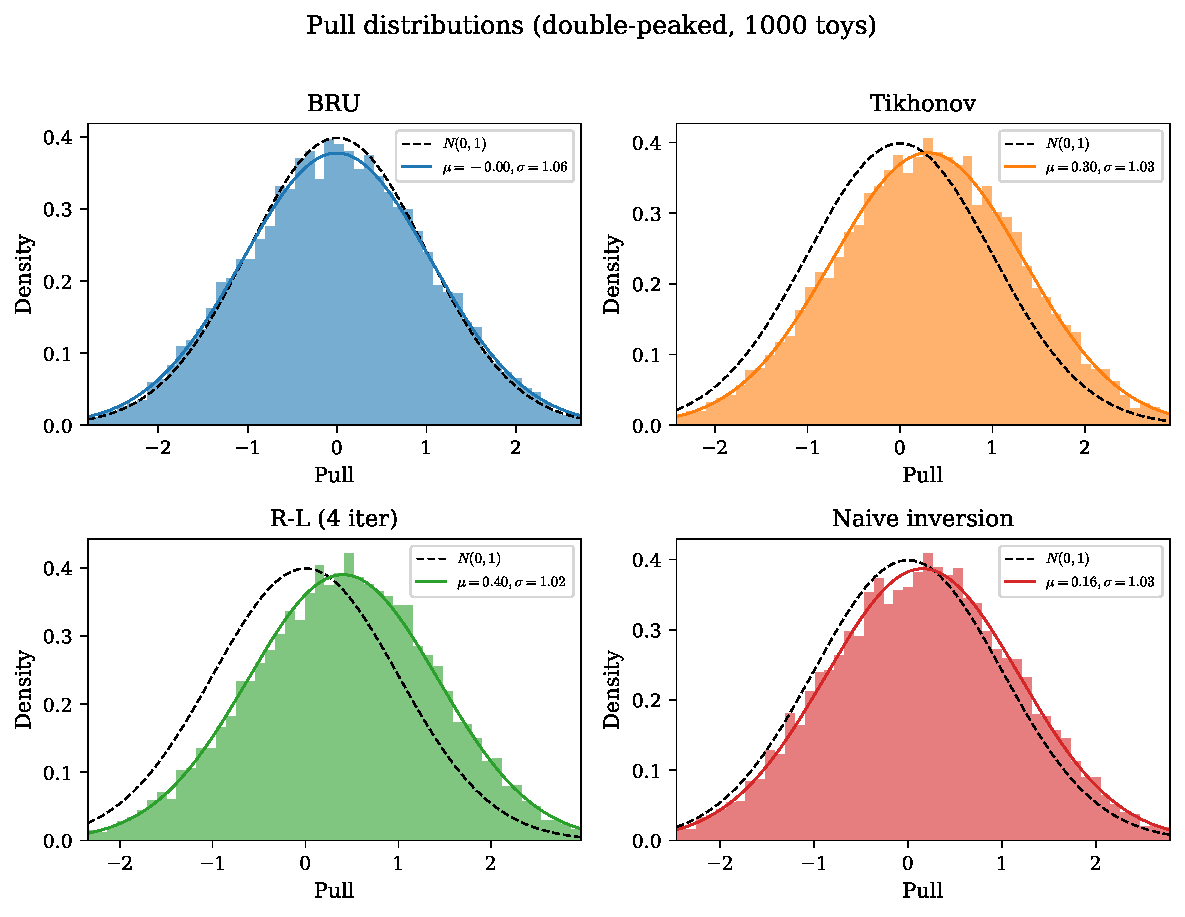}
  \caption{Pull distributions from 1,000 pseudo-experiments for the 
    double-peaked distribution.  The dashed curve shows the standard 
    normal $N(0,1)$; the solid curve is a Gaussian fit to the 
    observed pulls.  The fitted mean and width are indicated in each 
    panel.}
  \label{fig:pulls_dp}
\end{figure}

Figure~\ref{fig:pulls_dp} shows the pull distributions for all four 
methods.  The \bru{} pulls have mean $\mu = 0.00$ and width 
$\sigma = 1.06$, in good agreement with the standard normal.  
Tikhonov pulls show a positive bias ($\mu = 0.30$) and slightly 
inflated width ($\sigma = 1.04$), indicating that the error estimates 
do not fully account for the regularization-induced bias.  
Richardson-Lucy pulls exhibit a positive shift ($\mu = 0.40$) 
and width $\sigma = 1.02$, and naive inversion produces 
pulls with $\sigma = 1.03$, reflecting the large statistical 
uncertainties rather than genuine bias.

Table~\ref{tab:summary_dp} summarizes the performance metrics.  
\bru{} achieves the lowest mean-squared error (0.0025) among all methods, roughly twice as good as Tikhonov (0.0046) and eight times 
better than naive inversion (0.0205).
The combination of nearly unbiased pulls, near-nominal coverage, and low MSE demonstrates that the automatic eigenmode-based regularization successfully identifies the appropriate smoothing level without manual intervention.
\bru{} achieves a mean analytical coverage of 67\%, consistent with the nominal level.
Tikhonov reaches 64\% coverage, Richardson-Lucy 62\%, and naive inversion 66\%.
The undercoverage of Tikhonov and Richardson-Lucy reflects the positive bias introduced by over-regularization or premature truncation, which is not accounted for in the reported uncertainties.

\begin{table}[htb]
  \centering
  \caption{Performance summary for the double-peaked distribution 
    (1,000 pseudo-experiments, 8,000 events each).}
  \label{tab:summary_dp}
  \begin{tabular}{lcccc}
    \toprule
    Method & Pull $\mu$ & Pull $\sigma$ & Coverage & MSE \\
    \midrule
    \bru{} & $0.00$ & $1.06$ & $0.67$ & $0.0025$ \\
    Tikhonov & $0.30$ & $1.04$ & $0.64$ & $0.0046$ \\
    R-L (4 iter) & $0.40$ & $1.02$ & $0.62$ & $0.0032$ \\
    Naive inversion & $0.16$ & $1.03$ & $0.66$ & $0.0205$ \\
    \bottomrule
  \end{tabular}
\end{table}

\subsection{Steeply falling distribution}
\label{sec:steeply_falling}

The second test distribution is
\begin{equation}
  f(x) \propto 5.0\, (1 - x)^4 + 0.8\, 
  e^{-\frac{1}{2}\left(\frac{x - 0.25}{0.06}\right)^2} \,,
  \label{eq:steeply_falling}
\end{equation}
normalized to unit integral.
This distribution presents a qualitatively different challenge: the dynamic range exceeds a factor of ten between $x = 0$ and $x = 1$, and the narrow bump at $x = 0.25$ sits on the steep slope of the power-law component.  
Such configurations are common in particle physics, where transverse momentum spectra, jet mass distributions, and invariant mass spectra typically fall steeply.

\begin{figure}[htb]
  \centering
  \includegraphics[width=0.75\textwidth]{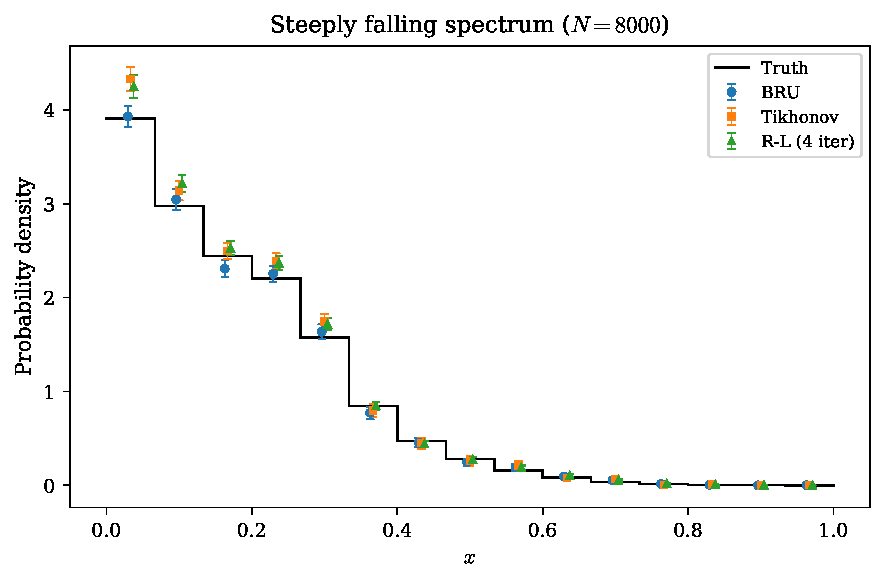}
  \caption{Single pseudo-experiment for the steeply falling 
    distribution.  Points with error bars show the unfolded results 
    from BRU, Tikhonov, and Richardson-Lucy; the solid histogram 
    is the truth.}
  \label{fig:singletoy_sf}
\end{figure}

Figure~\ref{fig:singletoy_sf} shows a single pseudo-experiment for 
this distribution.  \bru{} recovers the steeply falling shape 
faithfully, tracking the bump at $x = 0.25$ without introducing 
spurious oscillations in the tail.  Tikhonov regularization produces 
a somewhat smoother result that slightly underestimates the height 
of the bump.  Richardson-Lucy iteration, with only 4 iterations, 
struggles to converge in the rapidly changing region near $x = 0$ 
and shows visible distortions.

Table \ref{tab:summary_sf} presents the full summary statistics
for the steeply falling distribution.
\bru{} produces pulls with near-zero mean ($\mu = +0.01$) and a width $\sigma = 0.64$, indicating that the reported uncertainties are somewhat conservative in magnitude relative to the actual scatter, possibly signalling that the spline basis overestimates the variance in the tail where the curvature penalty dominates.
Because the bias-to-standard-error ratio is near zero in each bin, the analytical coverage of 68\% is consistent with nominal: the Kuusela formula (Eq.~\ref{eq:coverage}) depends on the mean pull per bin, not on its spread across toys.
Tikhonov shows a significant positive bias in the pulls ($\mu = 0.63$, $\sigma = 1.26$) and correspondingly reduced coverage (52\%).
Richardson-Lucy exhibits catastrophic behaviour on this distribution, as the Jacobian-propagated uncertainty estimate for R-L becomes degenerate in low-statistics bins where the algorithm has not converged, producing an error estimate of near zero that leads to a few pulls that dominate the uncertainty calculation by several orders of magnitude.
Naive inversion produces relatively well-behaved pulls ($\mu = 0.33$, $\sigma = 1.09$) but at the cost of very large uncertainties.

The steeply falling distribution, despite being relatively common distribution of particle spectra, exposes a fundamental limitation of several of algorithms.
For distributions with large dynamic range, the convergence rate differs dramatically between the high-statistics region near 
$x = 0$ and the low-statistics tail near $x = 1$.
Any given regularisation strength may be sufficient for the former but grossly inadequate for the latter, leading to systematic distortions that are not captured by any simple error estimate.
The \bru{} approach avoids this difficulty entirely because the regularization operates on each eigenmode independently, adapting to the signal-to-noise ratio in each component rather than applying a uniform convergence criterion across the entire distribution.

\begin{table}[htb]
  \centering
  \caption{Performance summary for the steeply falling distribution 
    (1,000 pseudo-experiments, 8,000 events each). Note that the pull estimates for R-L demonstrate the numerical failure of the uncertainty propagation in some bins as described in the text.}
  \label{tab:summary_sf}
  \begin{tabular}{lcccc}
    \toprule
    Method & Pull $\mu$ & Pull $\sigma$ & Coverage & MSE \\
    \midrule
    \bru{} & $0.01$ & $0.64$ & $0.68$ & $0.076$ \\
    Tikhonov & $0.63$ & $1.26$ & $0.52$ & $0.016$ \\
    R-L (4 iter) & - & - & $0.41$ & $0.014$ \\
    Naive inversion & $0.33$ & $1.09$ & $0.62$ & $0.033$ \\
    \bottomrule
  \end{tabular}
\end{table}

% =====================================================================
\section{Discussion}
\label{sec:discussion}
% =====================================================================

The numerical studies demonstrate that \bru{} achieves three desirable properties simultaneously.
The pull distributions are centered near zero on both benchmark distributions, indicating that the point estimates are unbiased.  The analytical per-bin coverage is close to the nominal 68\% level on both distributions, providing the statistical guarantees required for hypothesis testing and confidence interval construction.
The mean-squared error is competitive with or better than the alternatives, showing that the regularization does not sacrifice accuracy for the sake of statistical calibration.

The source of these advantages can be traced to two features of the 
method.
The B-spline representation ensures that the unfolded distribution is inherently smooth, eliminating the unphysical bin-to-bin fluctuations that plague histogram-based methods even 
after regularization.
The eigenmode decomposition provides a principled basis for the regularization, because the signal-to-noise ratio can be assessed independently for each mode rather than through a single global parameter that must balance competing requirements across the entire distribution.

The analytical per-bin coverage of \bru{} is consistent with the 
nominal 68.3\% level, confirming that the error propagation through 
the normalized spline representation is correctly calibrated.
The key requirement is that the Jacobian of the density estimate with respect to the spline coefficients include the normalization correction term, which accounts for the constraint that the estimated distribution integrates to unity; omitting this term inflates the reported errors and produces artificial overcoverage.

The comparison with Tikhonov regularization is particularly 
instructive.
Both methods employ a curvature penalty, but they differ in how the penalty interacts with the parametrization.
Tikhonov regularization, as shown by Kuusela and Panaretos~\cite{kuusela2015}, introduces a bias towards the regularization prior that the reported uncertainties do not account for, producing undercoverage that worsens as the regularization strength increases.
The observed reduction from the nominal 68\% to 64\% in our Tikhonov comparison is consistent with this finding.
In \bru{}, the curvature penalty is defined on the continuous spline function and is therefore independent of the evaluation binning, which is applied only at the final stage when presenting results.

The Richardson-Lucy algorithm, while elegant in its simplicity and positivity guarantee, lacks a natural mechanism for uncertainty estimation.
The number of iterations serves as an implicit regularization parameter, but there is no corresponding covariance matrix that accounts for the bias introduced by early stopping.
The catastrophic behaviour observed on the steeply falling distribution underscores the sensitivity of the method to the choice of iteration count, which may need to be tuned separately for different regions of the distribution.

% =====================================================================
\section{Representational fidelity and inferential transparency}
\label{sec:interpretation}
% =====================================================================

Unfolding is a means to an end.
The unfolded histogram is not itself the measurement; it is an intermediate representation whose purpose is to enable subsequent inference, whether that takes the form of fitting a theoretical model, testing a hypothesis, or comparing results from different experiments.
This distinction matters because the statistical properties of a regularized unfolding result are not those of a direct measurement, and treating the unfolded bin contents as if they were independent Gaussian observations with diagonal uncertainties will, in general, produce incorrect inferences.

The issue was identified clearly by Choudalakis in the context of Fully Bayesian Unfolding~\cite{choudalakis2012}, where the analogous warning concerns projecting the full posterior onto marginal histograms.
For classical regularized methods the situation is no different in principle: the regularization penalty couples neighbouring spline coefficients (or histogram bins, in Tikhonov-based approaches), inducing off-diagonal correlations in the unfolded covariance matrix.
A chi-squared comparison between the unfolded result and a theory prediction that ignores these correlations will overestimate the number of statistically independent degrees of freedom and may attribute genuine tension to spurious fluctuations or, conversely, mask real disagreement.
The correct procedure is to use the full covariance matrix $\hat{\mat{\Sigma}}$ propagated through the unfolding transformation, or equivalently to perform inference directly in the eigenmode basis where the covariance is diagonal by construction and each mode's contribution to any chi-squared is correctly weighted.

\bru{} is particularly well suited to this demand because the eigenmode decomposition is not an afterthought appended to the result but the core of the algorithm.
The filter factors $h_k$ make explicit, for every mode, how much of the unfolded amplitude reflects the data and how much reflects the smoothness prior.
A downstream fit to theory can therefore be performed in the eigenmode basis, retaining only those modes for which $h_k$ is appreciably different from zero, with the assurance 
that the reported covariance in that basis is diagonal and correctly 
normalized.
This is a materially stronger statement than what other regularized methods provide: it is not merely that the covariance is available, but that the method produces a natural orthogonal decomposition in which the inferential structure is transparent.

Recent developments in machine-learning-based unfolding, most prominently OmniFold~\cite{andreassen2020}, have framed the problem in terms of representational fidelity: the stated goal is to recover the full truth-level distribution without the binning artifacts that constrain histogram-based methods, using the flexibility of neural networks as a virtue.
This is a legitimate concern, and the push toward unbinned methods reflects a real limitation of coarsely binned unfolding.
However, representational fidelity and inferential transparency are not the same objective, and optimizing for the former can come at the expense of the latter.

In OmniFold, as an example, the regularization is implicit: it resides in the network architecture, the loss function, the number of training epochs, and the choice of early stopping criterion.
There is no decomposition analogous to the eigenmode expansion that would identify which components of the unfolded distribution are data-driven and which reflect the inductive bias of the model.
A physicist wishing to test whether the unfolded distribution is consistent with a theoretical prediction faces the problem of quantifying, in frequentist terms, a confidence interval on a reweighted event sample whose statistical properties depend on training details that are not part of the physics model.

The comparison illuminates what it means to provide an interpretable frequentist estimate of a distribution.
\bru{} sacrifices some representational flexibility in that the B-spline basis is smoother than a sufficiently expressive neural network, but it retains full accountability for the inferential consequences of that smoothness.
The signal-to-noise ratio of each mode is known, the bias introduced by regularization is quantifiable, and the coverage properties are accessible through toy Monte Carlo studies of the kind presented in Section~\ref{sec:numerical}.
These are the properties that matter when the unfolded result is to be used as input to a physics measurement, where frequentist guarantees on confidence intervals are not optional.

The appropriate criterion for evaluating an unfolding method is therefore not solely how faithfully it can represent an arbitrary truth-level distribution, but how clearly it communicates what was learned from the data and with what statistical warrant.
On this criterion, methods that expose their inferential structure, as \bru{} does through its eigenmode decomposition, have an advantage that cannot be recovered by post-hoc uncertainty estimation on a more flexible but opaque estimator.

% =====================================================================
\section{Conclusions}
\label{sec:conclusions}
% =====================================================================

We have presented a detailed exposition of Blobel's Regularized Unfolding, a method that combines cubic B-spline representations with eigenmode-based regularization to solve the inverse problem arising in particle physics measurements.
The method determines the regularization strength automatically from an internal consistency condition, eliminating the need for manual tuning.

Numerical studies on two benchmark distributions demonstrate that \bru{} produces pull distributions close to the standard normal, achieves per-bin coverage consistent with the nominal 68\% level, and delivers the lowest mean-squared error among the four methods 
compared.
These results hold both for a double-peaked distribution that tests resolution of localized features and for a steeply falling spectrum that tests performance across a large dynamic 
range.

The mathematical structure of the method, particularly the eigenmode decomposition that allows independent assessment of each component's signal-to-noise ratio, provides both practical advantages and physical transparency.
Each eigenmode has a clear interpretation as a spatial frequency of the distribution, and the filter factors make explicit which components of the solution are determined by the data and which are constrained by the smoothness assumption.

A Python implementation of \bru{} is available as part of the 
\texttt{histimator}\cite{histimator} and \texttt{RooUnfold}\cite{adye2011} packages, providing a modern interface to Blobel's algorithm for use in current and future analyses at the Large Hadron Collider and other experiments. A standalone translation of the original Fortran implementation of \texttt{RUN} was used as the basis for these implementations and is available on request.

\section{Acknowledgements}
\label{sec:ack}
This work began for me in 2017, at the suggestion of Olaf Behnke who encouraged me to dive into many of Volker's fascinating projects. 
Volker's many informative documents, slides and other publications (including the legacy \texttt{RUN} code) make up the vast majority of this work, and I am grateful also to Stefan Schmitt for answering my plea to continue to host this excellent repository of knowledge. 
I would also like to thank my colleagues in the RooUnfold team, particularly Lydia Brenner, for their excellent comments in helping this document and the associated method it contains, finally return to a useable state. 

\printbibliography

\end{document}